\documentstyle[epsf]{mn}

\newif\ifAMStwofonts

\def\ion#1#2{#1$\;${\small\rm{#2}}\relax}
\def\lesssim{\la}

\def\Ha{H$\alpha$}
\def\ha{\rm H\alpha}
\def\hii{\ion{H}{II}}

\def\etal{et\thinspace al.~}
\def\eg{e.g.,~}

\def\msol{\cal\,M_\odot}
\def\ergs{{\rm\,erg\,s^{-1}}}

\ifoldfss
  \ifCUPmtlplainloaded \else
    \NewTextAlphabet{textbfit} {cmbxti10} {}
    \NewTextAlphabet{textbfss} {cmssbx10} {}
    \NewMathAlphabet{mathbfit} {cmbxti10} {} 
    \NewMathAlphabet{mathbfss} {cmssbx10} {} 
  \fi
  \ifAMStwofonts
    \ifCUPmtlplainloaded \else
      \NewSymbolFont{upmath} {eurm10}
      \NewSymbolFont{AMSa} {msam10}
      \NewMathSymbol{\upi}     {0}{upmath}{19}
      \NewMathSymbol{\umu}     {0}{upmath}{16}
      \NewMathSymbol{\upartial}{0}{upmath}{40}
      \NewMathSymbol{\leqslant}{3}{AMSa}{36}
      \NewMathSymbol{\geqslant}{3}{AMSa}{3E}

    \fi
  \fi
\fi 

\ifnfssone
  \newmathalphabet{\mathit}
  \addtoversion{normal}{\mathit}{cmr}{m}{it}
  \addtoversion{bold}{\mathit}{cmr}{bx}{it}
  \newmathalphabet{\mathbfit} 
  \addtoversion{normal}{\mathbfit}{cmr}{bx}{it}
  \addtoversion{bold}{\mathbfit}{cmr}{bx}{it}
  \newmathalphabet{\mathbfss} 
  \addtoversion{normal}{\mathbfss}{cmss}{bx}{n}
  \addtoversion{bold}{\mathbfss}{cmss}{bx}{n}
  \ifAMStwofonts
    \ifCUPmtlplainloaded \else
      \UseAMStwoboldmath
      \makeatletter
      \new@mathgroup\upmath@group
      \define@mathgroup\mv@normal\upmath@group{eur}{m}{n}
      \define@mathgroup\mv@bold\upmath@group{eur}{b}{n}
      \edef\UPM{\hexnumber\upmath@group}
      \new@mathgroup\amsa@group
      \define@mathgroup\mv@normal\amsa@group{msa}{m}{n}
      \define@mathgroup\mv@bold\amsa@group{msa}{m}{n}
      \edef\AMSa{\hexnumber\amsa@group}
      \makeatother
      \mathchardef\upi="0\UPM19
      \mathchardef\umu="0\UPM16
      \mathchardef\upartial="0\UPM40
      \mathchardef\leqslant="3\AMSa36
      \mathchardef\geqslant="3\AMSa3E
    \fi
  \fi
\fi 

\ifnfsstwo
  \DeclareMathAlphabet{\mathbfit}{OT1}{cmr}{bx}{it}
  \SetMathAlphabet\mathbfit{bold}{OT1}{cmr}{bx}{it}
  \DeclareMathAlphabet{\mathbfss}{OT1}{cmss}{bx}{n}
  \SetMathAlphabet\mathbfss{bold}{OT1}{cmss}{bx}{n}
  \ifAMStwofonts
    \ifCUPmtlplainloaded \else
      \DeclareSymbolFont{UPM}{U}{eur}{m}{n}
      \SetSymbolFont{UPM}{bold}{U}{eur}{b}{n}
      \DeclareSymbolFont{AMSa}{U}{msa}{m}{n}
      \DeclareMathSymbol{\upi}{0}{UPM}{"19}
      \DeclareMathSymbol{\umu}{0}{UPM}{"16}
      \DeclareMathSymbol{\upartial}{0}{UPM}{"40}
      \DeclareMathSymbol{\leqslant}{3}{AMSa}{"36}
      \DeclareMathSymbol{\geqslant}{3}{AMSa}{"3E}
    \fi
  \fi
\fi 

\ifCUPmtlplainloaded \else
  \ifAMStwofonts \else 
    \def\upi{\pi}
    \def\umu{\mu}
    \def\upartial{\partial}
  \fi
\fi

\title[H\,{\normalsize\it II} regions and stellar ionizing sources]
	{
	Comparison of H\,{\Large\bf II} region luminosities with observed \\ 
	stellar ionizing sources in the Large Magellanic Cloud }
\author[M. S. Oey and R. C. Kennicutt, Jr.]
       {M. S. Oey$^1$\thanks{Email:  oey@ast.cam.ac.uk (MSO); 
	robk@as.arizona.edu (RCK)}
	and R. C. Kennicutt, Jr.$^{2\star}$\\
	$^1$Institute of Astronomy, University of Cambridge, Madingley Road,
	Cambridge   CB3 0HA, U.K. \\
	$^2$Steward Observatory, University of Arizona, Tucson, AZ\ \ 85721,
	U.S.A.}
        
\date{
Accepted 1997 August 12.
      Received 1997 July 31;
      in original form 1997 May 7.}

\pagerange{\pageref{firstpage}--\pageref{lastpage}}
\pubyear{1997}

\begin{document}

\maketitle

\label{firstpage}

\begin{abstract}

We estimate the total predicted Lyc emission rates of
OB associations for which the complete census of O star
spectral types exists.  The results are compared to the observed
\Ha\ luminosities of the host \hii\ regions.  We find evidence for
substantial leakage of ionizing photons from some \hii\ regions, while
others appear to be radiation bounded.  We estimate that overall for
the LMC, 0--51\% of the ionizing radiation escapes the local 
nebulae, and would be available to ionize the diffuse, warm,
ionized medium (WIM) in that galaxy.  This range of
values is consistent with the observed 35\% fraction of 
\Ha\ luminosity emitted by the WIM in the LMC, as well as
the corresponding fractions observed in other nearby galaxies.
It is therefore possible that photoionization by O stars is indeed
the dominant ionization mechanism for the WIM.

\end{abstract}


\section{Introduction}

The balance of evidence currently suggests that the ionization of the
diffuse, warm, ionized component of the interstellar medium (ISM)
is caused primarily by O stars.  From an energetic standpoint, this
stellar population has long been targeted as one of the only sources
capable of generating the large power requirement (\eg Reynolds 1984)
of this warm, ionized medium (WIM).  Models of the WIM that assume
ionization by O stars are broadly consistent with its observed
properties in the Galaxy (Miller \& Cox 1993; Dove \& Shull 1994), and
confirm that O stars are more than capable of providing 
the necessary ionizing power.  In fact, these studies suggest
that an excess of ionizing luminosity is produced, implying the
escape of that radiation from the Galaxy.  Observations of the WIM in
nearby, external galaxies show concentrations of the diffuse gas near
conventional \hii\ regions, further suggesting the association of the
WIM with O stars (Walterbos \& Braun 1994; Ferguson {\etal}1996).

However, there are notable complications to the O star ionization
hypothesis.  Perhaps most problematic is the observed limit to the 
ratio of the recombination lines \ion{He}{I} $\lambda5876$ / \Ha, implying
relative ionizing photon emission rates for He vs. H of $Q({\rm
He^0})/Q({\rm H^0}) \lesssim 0.03$  (Reynolds \& Tufte
1995; Heiles {\etal}1996).  The implied ionizing spectrum in the
Galaxy therefore appears to be much softer than anticipated from the O
star population, implying an unusually low effective upper-mass cutoff
to the stellar 
initial mass function (IMF) of $\lesssim 30 \msol$, and forcing an
unrealistically large Galactic star formation rate (Heiles {\etal}1996).
Rand (1997) presents a deep spectrum of the WIM in the edge-on galaxy
NGC 891, finding $Q({\rm He^0})/Q({\rm H^0}) \sim 0.08$, which is much
more consistent with hot star ionization.  Measurements of the He and
H recombination lines in the WIM of
three dwarf irregular galaxies by Martin \& Kennicutt (1997) are
also consistent with He being mostly ionized in those objects, as expected
from phoionization of stars with a normal IMF.
However, NGC 891 exhibits an anomalously high ratio 
of [\ion{N}{II}]$\lambda6583$/\Ha\ $\sim 1 - 1.4$, again implying an even
harder ionizing spectrum than indicated by Rand's \ion{He}{I}/\Ha\ ratio.
There is also some doubt about the ability of the stellar ionizing
radiation to travel the required hundreds of parsecs,
although models by Dove \& Shull (1994) and Miller \& Cox (1993) show
tentative compatibility.  Investigations of the superbubble
structure of the ISM (Heiles 1990; Rosen \& Bregman 1995; Oey \&
Clarke 1997) also suggest the 
widespread existence of large voids, which would facilitate radiative
transfer over such large distances.  Finally, alternative or additional
ionizing sources are likely to play a role in the WIM
as well.  These include turbulent mixing layers (Slavin, Shull, \&
Begelman 1993), neutrino decay from dark matter (Sciama 1990,
1995), white dwarfs, cosmic rays (Liu
\& Dalgarno 1997), and extragalactic ionizing radiation (e.g., Reynolds
{\etal}1995). 

In general, however, O stars remain as the most likely dominant source
of ionization for the WIM.  But is the simple, empirical comparison of
available ionizing photons vs. ionized gas actually consistent with
this scenario?  Abbott (1982) predicts that roughly
15\% of the available ionizing flux of O stars 
in the solar neighborhood is required to ionize the local WIM, a result
supported by Dove \& Shull (1994) and Miller \& Cox (1993).  This
therefore implies that, if the large-scale WIM is similar
to that in the solar neighborhood, at least $\sim15$\% of the OB Lyman
continuum (Lyc) 
flux must escape the \hii\ regions.  Photoionization models of
the WIM by Domg\"orgen \& Mathis (1994) further suggest that 4\% of
the total flux must escape the Galaxy altogether, in which case a
certain fraction of the ionized regions must be density-bounded.
However, historically, most \hii\ regions have been considered to be
essentially radiation-bounded, rather than density-bounded, allowing
the nebular luminosities to be used quite successfully
as tracers of massive star formation.  It is
therefore worth examining more closely the comparison of available Lyc and
nebular fluxes.

We can investigate this effect with a sample of OB associations
whose stellar populations have been classified, and whose host nebulae
have measured total luminosities.  The Large Magellanic Cloud (LMC)
provides an ideal sample, with detailed studies of over a dozen OB
associations (\eg Massey {\etal}1995b; Oey 1996a), and a uniform catalog
of nebular photometry (Kennicutt \& Hodge 1986) at low extinction.  We
now use this 
sample of OB/\hii\ systems to examine the fraction of Lyc radiation
escaping from the \hii\ regions, as indicated by the currently
available stellar atmosphere models.

\section{Methods}

Table~\ref{sample} lists the sample of OB associations, designated
by their Lucke \& Hodge (1970) identification; and host nebulae,
given by their Davies, Elliott, \& Meaburn (1976), and
Henize (1956) identifications.  The OB associations all have available
spectroscopic classifications for the hottest stars.  These
classifications are essential in constraining
the Lyc fluxes of the stars, since the spectral types cannot be
reliably distinguished from broad-band colors alone (e.g., Massey 1985).
The reference for the spectral types is given in the
fourth column of Table~\ref{sample}.

\begin{table}
\centering
\caption{Sample of OB / \hii\ Systems \label{sample}}
\begin{tabular}{@{}llll}

LH & DEM & Henize & Reference \\

2      & 10B & N79 CE & J. Wm. Parker, unpublished \\
... & 25 & N185 & Oey (1996a) \\
6       & 31 & N9   & Oey (1996a) \\
9, 10  & 34  & N11  & Parker {\etal}(1991) \\
... & 50 & N186 & Oey (1996a) \\
38     & 106 & N30 BC & Oey (1996a) \\
47, 48 & 152, 156$^a$ & N44 & Oey \& Massey (1995) \\
51, 54 & 192 & N51 D & Oey, in preparation \\
58     & 199 & N144 & Garmany {\etal}(1994) \\
73     & 226 & N148 I & Oey (1996a) \\
83     & 243 & N63 A & Oey (1996a) \\
110    & 293 & N214 C & Conti {\etal}(1986) \\
114    & 301 & N70  & Oey (1996a) \\
117, 118 & 323, 326 & N180 & Massey {\etal}(1989) \\
\end{tabular}

\medskip

$^a$Not including DEM 152A. \\
\end{table}

To compare the predicted stellar ionizing flux with the \hii\ region
luminosity, we summed the Lyc photon emission rates $Q({\rm H}^0)$ for the
individual stars, that have been
estimated for the range of O star spectral types.  Although this has
already been carried out for several of the nebulae in this sample, we
now redo these comparisons with more recent information and uniformly
for the entire sample.  At present, the
values of $Q({\rm H}^0)$ are rather uncertain, as discussed below.  We
therefore compute results using the tabulation by Panagia (1973,
hereafter P73), Vacca, Garmany, \& Shull (1996, hereafter VGS), and
Schaerer \& de Koter (1997, hereafter SdK).  We then computed the
resulting \Ha\
luminosity implied by the total $Q({\rm H}^0)$ of each OB association, assuming
an electron temperature $T_e = 10,000$K.  We reconfirmed the Lyc/\Ha\ 
photon conversion of 2.2 for $T_e=10,000$K from the results of Hummer
\& Storey (1987).  The stellar studies also provide extinction data
with which we arrive at a final estimate for the predicted \hii\
region luminosity, using $A_{\ha} = 2.5\ E(B-V)$ (\eg Parker
{\etal}1992).  The adopted $E(B-V)$ are median color excesses, with the
exception of values for DEM 34, DEM 199, and DEM 323/326, which are
mean values.  The adoption of the median or mean is taken directly
from the cited reference; differences in the respective
characterization of $E(B-V)$ have negligible effect in this analysis.
Table~\ref{compare} lists the nebular identification in
column 1, and in columns 3 -- 5, the extinction-corrected \Ha\
luminosities predicted from the spectral type -- Lyc flux conversions of 
P73 ($L_{\rm P}$), VGS ($L_{\rm VGS}$), and SdK ($L_{\rm SdK}$).
Column 6 shows the total $Q({\rm H}^0)$ for the classified O stars, using the
SdK conversion; the different conversions and choice among these are
discussed in \S 3 below.  The number of classified O stars 
($n_\star$) and $B-V$ color excess are given in columns 7 and 8.

\begin{table*}
\centering
\begin{minipage}{140mm}
\caption{Observed and Predicted \hii\ Region Luminosities \label{compare}}
\begin{tabular}{@{}lcccccrcc@{}}
\hii\ Region & $L_{\rm obs}$ & $L_{\rm P}$ & $L_{\rm VGS}$ & 
	$L_{\rm SdK}$ & $Q({\rm H}^0)$ & $n_\star$ & $E(B-V)$ & 
	$L_{\rm obs}/L_{\rm SdK}$  \\
& ($\ergs$) & ($\ergs$) & ($\ergs$) & ($\ergs$) & $({\rm s}^{-1})$ & & 
	(mag) & \\
\\
DEM 10B   &  6.68E+37  & 5.33E+37 & 9.61E+37 & 8.24E+37 & 8.64E+49 &  7
	& 0.16 & 0.81 \\
DEM 25 &     2.64E+37  & 2.23E+36 & 3.88E+36 & 3.08E+36 & 2.88E+48 & 1
	& 0.11 & 8.57  \\
DEM 31 &     6.42E+37  & 1.11E+38 & 1.73E+38 & 1.62E+38 & 1.45E+50 & 6
	& 0.09 & 0.40  \\
DEM 34 &  5.46E+38  & 6.83E+38 & 9.37E+38 & 8.33E+38 & 7.61E+50 & 
	44$^*$ & 0.10 & 0.66  \\
DEM 50    &  4.61E+37  & 1.61E+37 & 2.70E+37 & 2.30E+37 & 2.20E+49 & 3
	& 0.12 & 2.00  \\
DEM 106 &    3.43E+37  & 4.15E+37 & 7.34E+37 & 5.55E+37 & 5.56E+49 & 8
	& 0.14 & 0.61  \\
DEM 152+156$^a$ & 2.32E+38  & 3.18E+38 & 4.11E+38
	& 3.52E+38 & 3.22E+50 & 35$^*$ & 0.10 & 0.66  \\
DEM 192    & 2.50E+38  & 2.61E+38 & 3.66E+38 & 3.03E+38 & 2.71E+50 & 
	25$^*$ & 0.09 & 0.83  \\
DEM 199 & 4.09E+38  & 3.00E+38 & 3.94E+38 & 3.34E+38 & 2.72E+50 & 
	22$^*$ & 0.05 & 1.22  \\
DEM 226    & 2.23E+37  & 1.65E+37 & 2.85E+37 & 2.41E+37 & 2.53E+49 & 4
	& 0.16 & 0.93  \\
DEM 243    & 5.22E+37  & 6.14E+37 & 1.19E+38 & 9.72E+37 & 8.88E+49 & 11
	& 0.10  & 0.54 \\
DEM 293 &    4.97E+37  & 8.11E+37 & 4.78E+37 & 4.56E+37 & 4.79E+49 &  1
	& 0.16 & 1.09  \\
DEM 301 &    4.84E+37  & 1.87E+38 & 2.65E+38 & 2.45E+38 & 2.04E+50 & 7
	& 0.06 & 0.20  \\
DEM 323+326 & 3.30E+38 & 3.88E+38 & 3.24E+38 & 2.92E+38 & 2.80E+50 & 20
	& 0.12 & 1.13  \\
\end{tabular}

\medskip
$^*$WR star excluded; DEM 199 contains 3 WR stars. \\
\smallskip
$^a$Not including DEM 152A.

\end{minipage}
\end{table*}

The \Ha\ luminosities of the \hii\ regions were measured using the 
photographic photometry of Kennicutt \& Hodge (1986, hereafter KH86).
Details of the observations and data can be found in the original
paper.  The plates used in KH86 were rescanned at higher angular
resolution and recalibrated using sensitometer exposures obtained
during the observations, photoelectric photometry from KH86 and Caplan
\& Deharveng (1986), and CCD images of 30 Doradus obtained by RCK with
the CTIO 1.5m telescope and Rutgers Fabry-Perot camera.  This produced
a series of \Ha\ surface brightness maps corresponding to each of the
four photographic fields covering the LMC.  Integrated fluxes for each
\hii\ region were measured, using the stellar data to ensure that the
nebular regions measured conformed as closely as possible to the
regions ionized by the OB associations of interest.  The boundaries of
the \hii\ regions are well-defined in most instances, so the
determination of the apertures for the \Ha\ flux measurements was
unambiguous.  No corrections
for extinction were applied to these data, since extinction is taken
into account for the predicted \Ha\ luminosities in Table~\ref{compare}.
The fluxes we derive show excellent agreement with the measurements of
Caplan \& Deharveng (1986) for objects in common (see below).  The
observed \Ha\ luminosities $L_{\rm obs}$ of the \hii\ regions in our
sample are given in column 2 of Table~\ref{compare}. 

The last column of Table~\ref{compare} shows the ratio of observed to
predicted \Ha\ luminosity, $L_{\rm obs}/L_{\rm SdK}$.  If
$Q({\rm H}^0)$ is a reasonable estimate of the true ionizing photon emission,
then we can use $L_{\rm obs}/L_{\rm SdK}$ to obtain the
fraction of ionizing radiation escaping from the \hii\ region.  For
DEM 25 and DEM 50, these values are $\gg 1$.  Interestingly, the morphology of
these two nebulae is similar.  Both objects are shells that
show evidence of recent supernova activity (Oey 1996b), although their
$L_{\rm obs}\sim 10^{37}\ \ergs$ is one to two orders of magnitude brighter
than typical supernova remnants (SNRs).  However, simulations of
supernova impacts on pre-existing shells show that peak \Ha\
luminosities of $\sim 10^{37}\ \ergs$ might be achieved for a
short flash in time (Tenorio-Tagle {\etal} 1990).
It therefore seems possible that there may be a dominant
contribution from shock ionization in DEM 25, where the photometry 
suggests that identification of the O stars is complete (Oey 1996a).
DEM 50 has a confirmed SNR at the north end of the main shell, and it
is difficult to determine its contribution to the total \Ha\ luminosity.
However, photometry for this region (Oey 1996a) indicates that
significant O stars may remain unclassified for DEM 50 in
comparison to the other nebulae in the sample.  For these reasons, we will
not consider DEM 25 and DEM 50 any further in the comparison of Lyc
emission to \Ha\ luminosity.  We also caution that DEM 243 contains an
SNR as well, but given the ratio of $L_{\rm obs}/L_{\rm SdK}$, it is
apparent that the SNR does not contribute significantly to the \Ha\
luminosity. 

\section{Uncertainties}

The median of the ratio $L_{\rm obs}/L_{\rm SdK}$, the observed to
predicted \Ha\ luminosity, is 0.74, excluding DEM 25 and 50.
Because of the possibility of unidentified stars and other
effects yielding an asymmetric distribution, the median is the
preferred measurement of a characteristic value, although we note that
the mean value  is 0.76, in close agreement with the median.  There
are large uncertainties in the median estimate, stemming from many
factors.  We adopt 
a distance modulus to the LMC of 18.5 (\eg Panagia {\etal}1991), with
an error of $\pm 10$\%.  Our adopted nebular electron temperature
$T_e= 10,000$K (\eg Pagel {\etal}1978), yielding an
uncertainty in the ratio of \Ha\ photons to Lyc photons of $\pm 1$\%.
Additional important errors result from the effect of stars
that are unaccounted for, uncertainties in modeled Lyc emission rates,
and observational errors.

The number of classified O stars is a lower limit to the actual number
present, and the values of $Q({\rm H}^0)$ in Table~\ref{compare} do not include
contributions from B stars and Wolf-Rayet (WR) stars.  All of the WR
stars in these \hii\ regions are WNE or WC types.  DEM 34, DEM
152+156, and DEM 192 each contain one WR star, and DEM 199 contains 3
WR stars.  DEM 31 contains a WN6 that falls in the spectral sequence
of O3--4 If to WN-A stars (Oey 1996a), so we have attributed it the Lyc
emission of an O3 I star.  The ionizing fluxes of WR stars are quite
uncertain, and, as shown by modeling of individual stellar atmospheres
for WNL stars (Crowther \& Smith 1997), they can vary enormously for
individual spectral types.  Given this caveat, model atmospheres
suggest that WNE (Crowther \& Smith 1996) and WC types may be expected
to contribute roughly $10^{49}\ \rm s^{-1}$ to the cluster ionizing
emission (P. A. Crowther, private communication). 

Early B stars may provide an additional significant
source of ionization.  $EUVE$ observations of the B2 II star,
$\epsilon$ CMa, revealed excess Lyc emission by a factor of 30 over model
atmosphere predictions (Cassinelli {\etal}1995).  Subsequent
observations of the B1 II-III star, $\beta$ CMa, implied a possible
excess of up to a factor of 5, although by adjusting the stellar effective
temperature $T_{\rm eff}$, the observations could come into agreement with
predictions (Cassinelli {\etal}1996; SdK).  It is therefore unclear to what
degree $\epsilon$ CMa is an anomalous star, bearing in mind that it
was selected for observation based in part on its bright EUV flux.  It
is also interesting that the observations of these two B stars suggest
that the Lyc fluxes predicted by the models appear thus far to be
lower limits.  A discussion of effects and uncertainties in $Q({\rm H}^0)$
estimates for early B stars may be found in SdK.   

We also caution that predicted O star Lyc fluxes have not been
observationally confirmed.  As is apparent in Table~\ref{compare}, the
predictions of $Q({\rm H}^0)$ by P73 and VGS yield total ionizing luminosities
that agree within a factor of 2 or less, while those of SdK are generally 
intermediate.  The P73 values historically have been widely used, and
are based on LTE line-blanketed models for cooler spectral types, and
NLTE unblanketed models for hotter types.  VGS presented a new set of
$Q({\rm H}^0)$ estimates based on 
updated stellar parameters and using LTE line-blanketed models.  The
factor of $\lesssim 2$ difference between VGS and P73 is due primarily
to the updated $T_{\rm eff}$ calibrations.  Finally, the
recent results by SdK should represent the best estimates of Lyc
fluxes to date, since they account for NLTE, line-blanketing, and wind
effects, within self-consistent models combining both stellar
structure and atmospheres.  More detailed comparison and discussion of
relevant effects is carried out by VGS and SdK.  

All these models assume solar metallicity ($Z_\odot$).  SdK show that
a decrease in metallicity is expected to increase $Q({\rm H}^0)$ slightly,
owing to reduced line-blanketing.  The degree of the increase is
generally small, but a decrease from solar to $0.2 Z_\odot$ could
increase the Lyc output by up to 30\% for individual spectral types,
although these large effects apply primarily to the cooler 
types.  However, an empirical spectral type -- $T_{\rm eff}$
calibration has been lacking for low-metallicity stars, and if
temperature effects are important, these could well dominate the Lyc
output.  In general, though, most arguments suggest that lower
metallicity should tend to increase the stellar ionizing fluxes.
At any rate, with LMC abundances around $0.4 Z_\odot$,
the increase in Lyc fluxes should be relatively small in light
of the other uncertainties.  

Our reddening estimates (Table~\ref{compare}) based on the stellar
data are in good agreement with nebular extinctions measured by Caplan
\& Deharveng (1986) from the Balmer decrement.  For the
standard extinction assumptions described in their Appendix B, the
mean difference in $E\ (B-V)$ between their estimates and ours is 0.01
mag for the 7 objects measured in common.  The mean absolute value in
the differences is 0.03 mag, suggesting an uncertainty of $\pm 7$\%.

The principal sources of error in the \Ha\ measurements are spatial
variations in response across the Curtis Schmidt plates, and the
accuracy of the photographic density-intensity calibration, especially for low
surface brightness regions as are found in some of the superbubbles.
We estimate the typical uncertainties in the fluxes to be $\pm$10\%
for the brighter regions, increasing to $\pm$15\% in the superbubbles,
based on intercomparisons of objects observed on more than one plate,
and from comparisons with the photoelectric measurements described above.

The uncertainties in the median value of $L_{\rm
obs}/L_{\rm SdK} = 0.74$ that result from the factors mentioned above
are strongly dominated by the large uncertainty in model atmosphere
Lyc emission rates.  This error is difficult to quantify, but we
estimate that it is around 50\%.  We note that many effects such as
those due to line-blanketing, wind-blanketing, and metallicity, tend to
introduce uncertainties for cooler stars, whereas the \Ha\ luminosities
are dominated by the hotter stars.  
The effect of uncounted O stars, WR stars, and B stars suggests that
the error bars are actually asymmetric in favor of higher predicted 
Lyc emission, but these are still likely to be small effects in
comparison to that due to O star Lyc uncertainties.
Thus we estimate a median value of $L_{\rm obs}/L_{\rm SdK}$ in the
range 0.49 -- 1.1,
or equivalently, a mean fraction of Lyc radiation escaping from these
\hii\ regions in the range 0 -- 51\%.  Although formally the lower
limit to the range is 0\%, we note that the existence of individual nebulae
that are quite reliably density-bounded (Table~\ref{compare})
implies that the lower limit is $>0$, but is difficult to quantify further. 

\section{Discussion}

While the range of values suggests that a significant escape
fraction seems likely, it is also consistent with the nebulae being
generally radiation bounded, although Table~\ref{compare} shows that some
individual objects are convincingly density-bounded.  But we note that 
the sample of faint Galactic \hii\ regions studied by Hunter \& Massey
(1990) also showed an overall excess of available ionizing 
radiation.  That study predicted the Lyc emission rates from
observed, classified stellar populations using the P73 conversions, and
compared the results with the Lyc emission rates implied by \Ha\
and radio continuum observations.  Inspection of their results shows
that the median ratio of observed to
predicted emission rates is 0.7 for both the \Ha\ and radio-derived
comparisons, with 19 objects in the \Ha\ sample and 21 in the radio
sample.  The results from this Galactic study are thus in 
good agreement with the median value from our sample of 0.74.
Note that the use of the P73 Lyc predictions by Hunter \& Massey
(1990) imply that the median ratios from their sample are
therefore slight overestimates when comparing with our results (cf.
Table~\ref{compare}).

An escaping Lyc fraction of up to 51\% is 
fully consistent with the estimated fraction of diffuse to total \Ha\
flux of $35\pm 5$\% estimated by Kennicutt {\etal}(1995) for the LMC.  
It is therefore indeed quite possible that the WIM, at least in the LMC, is
ionized by hot supergiants.  If the \hii\ regions in the LMC are
typical in their structure, extinction, and relationship to the
diffuse ISM, then these results may be compared to observations of the
WIM in other galaxies.  Measurements for \Ha\ luminosity fractions of
the WIM in nearby galaxies range from $\sim20$\% to 53\% (see Ferguson
{\etal}1996), which are also consistent with the
fraction of ionizing radiation escaping from the LMC \hii\ regions.
Thus it is a likely possibility that the WIM in these galaxies is
dominated by O star photoionization.  With better constraints on the
large uncertainties in our estimate of the median $L_{\rm obs}/L_{\rm SdK}$, 
it should become possible to better evaluate this hypothesis, as well as
the role of alternate ionizing mechanisms, which
can contribute to the \Ha\ luminosities in both normal
\hii\ regions and the WIM as well.

\begin{figure*}
\vspace*{-1.9 truein}
\begin{center}
   \hspace*{1.0 truein}
   \epsfbox{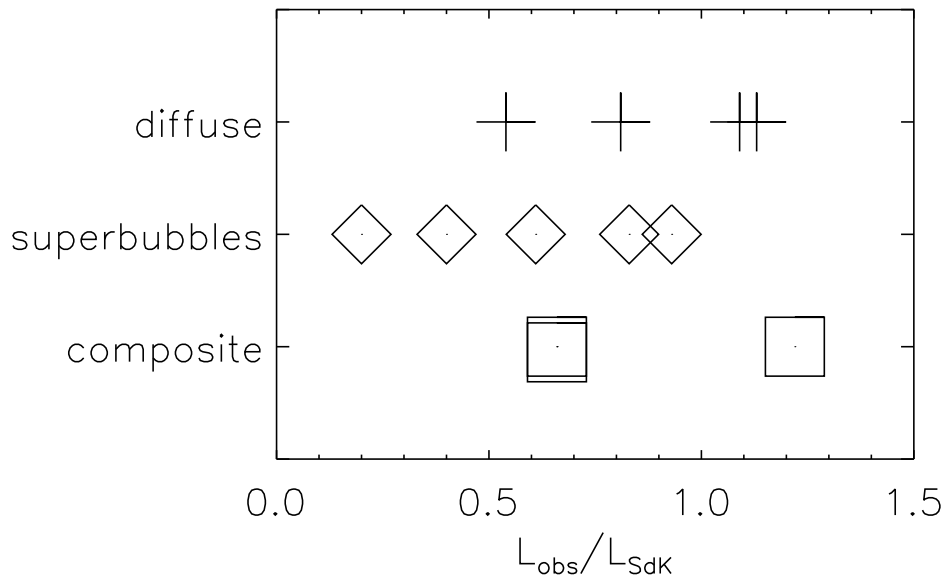}
\end{center}
\vspace*{-15pt}
\caption[]{Distribution of $L_{\rm obs}/L_{\rm SdK}$ among the different
nebular morphologies.  Diffuse \hii\ regions are shown with plus
symbols, superbubbles with diamonds, and composite objects with
squares.  There are two points at $L_{\rm obs}/L_{\rm SdK} = 0.66$ for
the composite objects. 
\label{morph}}
\end{figure*}

Our selection of \hii\ regions encompasses a range of nebular
morphology, including superbubbles, diffuse \hii\ regions, and
composite objects.  The distributions of $L_{\rm obs}/L_{\rm SdK}$ for
these three subsamples are compared in Figure~\ref{morph}.  On the whole,
the superbubbles may have lower ratios of observed to predicted \Ha\
luminosities; the mean for the 5 superbubbles in this sample is 0.59,
as compared to a mean of 0.89 for the 4 diffuse objects.  The mean for
the 3 composite objects is 0.85.  
A preferential leakage of ionizing radiation for the superbubbles might be
real if actual holes are present in some of the shell walls, as may be
suggested by the morphology of some objects.  Such holes could allow
the escape of ionizing radiation from those superbubbles.  
Overall, however, Figure~\ref{morph} shows that the differing 
morphologies all span a large range in $L_{\rm obs}/L_{\rm SdK}$.
In view of possible selection effects and small number statistics,
we hesitate at present to attribute any significance to the possible trend
with morphology. 

Our results are suggestive that many \hii\ regions may
be density bounded rather than ionization bounded.  As can be
seen in Table~\ref{compare}, this may not be the case for any
individual object, but appears likely for many.
Density-bounding of the nebulae will probably
have a noticeable effect in the observed emission of lower-ionization
species such as [\ion{O}{II}], [\ion{N}{II}], and [\ion{S}{II}], that
normally dominate in the outer 
regions of ionization-bounded nebulae (see, \eg Shields 1990).  
Emission-line ratios that are based on species originating in different 
nebular volumes should therefore be interpreted with some caution, if only
ionization-bounded models are considered.  Photoionization modeling is
necessary to fully investigate this effect, which will be explored in a
future study. 

The sample spans over one order of magnitude in \hii\ region
luminosity, over which there is no obvious correlation between
$L_{\rm obs}$ and fraction of escaping Lyc radiation.
Rozas, Beckman, \& Knapen (1996) have suggested that a change in slope
of the \hii\ region luminosity function around $10^{38.6} \ergs$ is
due to the density bounding of objects with greater luminosities,
whereas the fainter objects are suggested to be primarily ionization
bounded.  Most of the nebulae in our sample have luminosities below their
critical value, and as seen in Table~\ref{compare}, there is a
significant fraction of objects that may be convincingly identified
as density-bounded.  However, the LMC cannot provide a useful sample
of \hii\ regions with $\log L_{\rm obs} > 38.6$, hence it is impractical
to further test the hypothesis with this galaxy.


In summary, we find that the \hii/OB systems in the LMC suggest
that up to 51\% of the ionizing radiation from the hot stars is escaping
the local \hii\ regions.  A significant number of individual \hii\
regions reliably appear to be density-bounded.  At present, we find no
compelling correlation with nebular morphology or luminosity in the
fraction of Lyc photons escaping the \hii\ regions, although
superbubbles might possibly exhibit a greater escaping fraction.  Our
sample of objects should be fairly representative of the variety of
\hii\ regions in normal star-forming galaxies, as it includes
superbubbles, diffuse nebulae, and complex regions, all with varying
luminosities and galactic location.  The relative proportion of
different nebular types will ultimately depend on the star formation
history and environment in any given galaxy.
Our estimate of the Lyc escape fraction is consistent with the
observed fraction of total \Ha\ luminosity emitted by the WIM in the
LMC and other star-forming galaxies, and suggests that photoionization by
ordinary O stars could indeed be the dominant source of the ionization
for the WIM, although alternate ionization mechanisms are not ruled out.

\section*{Acknowledgments}

It is a pleasure to thank Paul Crowther, Daniel Schaerer, Keith Smith,
and Bill Vacca for discussions on stellar atmosphere models.  We also
benefited from useful comments received at the Boulder-Munich II
Workshop in Windsor, held in July, 1997.  We are grateful to
Joel Parker for access to spectral types for LH 2 in advance of
publication.  Thanks to the anonymous referee for a careful reading of
the manuscript.  This research was supported in part by the US National
Science Foundation through grant AST-9421145.

\label{lastpage}

\end{document}